\documentclass[twocolumn, a4paper]{article}
\usepackage{url}
\usepackage[]{algorithm2e}
\usepackage{amsmath}
\usepackage{float}
\usepackage{enumitem}
\usepackage{cite}

\restylefloat{table}
\begin{document}
\begin{center} {\Large Literature Review: Smart Contract Semantics}\end{center}
\begin{center}
\textbf{Varun Mathur}
\end{center}

\begin{abstract}
This review presents and evaluates various formalisms for the purpose
of modelling the semantics of financial derivatives contracts. The
formalism propsed by Lee\cite{lee1988} is selected as the best
candidate among those initially reviewed. Further examination and
evaluation of this formalism is done. 

\end{abstract}

\section{Introduction}
\subsection{Introduction to the Problem}
Contracts are legally binding documents that govern agreements between
some parties. Generally, they define obligations between parties,
typically regarding the delivery of goods, the performance of
services, or payments. Contracts are essential to business
transactions across all industries, as well as private transactions
amongst the general public. In this paper, the domain is limited
strictly to financial derivatives contracts that have been generated
from the International Swaps and Derivatives Association (ISDA) Master
Agreement template.

Contracts are a crucial part of doing business. Consequently,
businesses are heavily invested in the task of \textit{Contract
  Lifecycle Management} (CLM). CLM encapsulates all the various
processes and operations that go into supporting a contract throughout
its life including: contract creation, negotiation, appoval,
execution, and analysis. A study\cite{aberdeen2006} by the Aberdeen
Group found that 80\% of businesses were manually performing some or
all of these tasks. They found that standardising, formalising, and
ultimately automating these processes would result in cost savings
from greater efficiency. In their analysis of the global financial
derivatives industry, ISDA had similar findings
\cite{isdainfra2016}. Therefore, there is a clear need for a solution
that would allow us to automate some or all of these CLM processes.

Smart contracts are a solution to this problem. Clack et
al\cite{clack2016smart-foundations, DBLP:journals/corr/ClackBB16a}
offer this definition of a smart contract:

\textit{A smart contract is an automatable and enforceable
  agreement. Automatable by computer, although some parts may require
  human input and control. Enforceable either by legal enforcement of
  rights and obligations or via tamper-proof execution of computer
  code}

Smart contracts would allow us to automate and monitor the performance
of legal agreements electronically. For example, smart contracts could
standardise and automate the performance of actions that occur over
the lifetime of a financial derivative contract, reducing
infrastructure cost. To start with, only a small part of the legal
agreement would be automated, but as technology improves these
agreements would be increasingly automated, including automatic
detection of non-performance. Additionally, the use of smart contract
templates would simplify the task of generating complex legal
agreements between parties.

The first step in developing smart contracts, and the problem that
this paper examines, is formalisation. This is the conversion of a
contract into a formal representation that is unambiguous and machine
interpretable. This conversion must happen in a manner such that we
can be assured that the semantics of the contract and that of the the
formal representation are identical. Once a contract has been
converted to this form, we could write computer code to automate its
performance. This code could easily make reference to the features of
the contract as specified in its formal representation.

\subsection{Modelling the Semantics of Legal Agreements}
We adopt the terminology of Hvitved\cite{hvitved2011contract} when
discussing \textit{formalisms}. A formalism consists of a
\textit{formal model} and a \textit{formal language}. In this context,
a formal model is a mathematical model, the components of which can be
used to construct formal representations of contracts.  A formal
language is a syntactic representation of some model. There should
exist some mapping between the syntactic elements of the language and
the semantic elements of the model. Formalisms are necessary since
legal texts in their original form contain complex vocabulary and
language constructs (e.g.\ modal verbs, conditional futures, and
complex temporal clauses). This makes it difficult for a smart
contract engineer to write code to automate the performance of the
agreement directly from this form. Additionally, sometimes this
language can be ambiguous and difficult to parse. On the other hand,
the formal representation of this contract should be unambiguous and
machine interpretable by definition. Finally, by having a formal model
of the legal text, we can create automated validation tests for its
smart contract code.

There are various aspects of legal text that we would like to model
formally. An initial analysis yields three broad categories of things
we would like to model:

\paragraph{Deontic} These are aspects relating to rights, obligations,
permissions, and prohibitions. Typically, parties to a contract do not
spontaneously take action unless they have the right or an obligation
to do so as specified by the contract. If a party does not perform or
deliver on their obligations according the constraints specified in
the contract, they will typically be subject to some punitive
action. This may be specified in detail in the contract, or it may be
left to a court of law to decide and administer appropriate punishment
for non performance. When a party has not fulfilled its obligations or
takes some prohibited action, it is said to be in a state of
\textit{default}. The deontic relationships defined in contracts are
typically only ``active'' under some conditions. For example, in a
contract governing the exchange of goods the Buyer may only be
\textit{obligated} to pay the Seller after the goods have been
delivered. Another way to express this is that the Buyer's obligation
to pay is only \textit{activated} after the Seller has fulfilled their
obligation to deliver the goods.

\paragraph{Temporal}
The temporal aspects of a contract are aspects related to time.
Temporal expressions in legal agreements are typically used to specify
deadlines for obligations or the ``life spans'' of permissions and
prohibitions. These deadlines may be absolute temporal expressions
(e.g. \textit{Party A must deliver the goods by 12:00 on Friday 4
  November 2018}) or they may be relative temporal expressions
(e.g. \textit{Party B must pay Party A \pounds 400 within 7 days of
  receiving the goods.}) The ``life spans'' of permissions and
prohibitions typically take the form of continuous time intervals.

\paragraph{State, Events, and Actions}
As discussed, contracts have some notion of state, since their
semantics may change over time as parties make choices and perform
actions. While the operational aspects of a contract covers
state-modifying actions that are defined in the contract, we must also
consider how changes in external state and actions that parties take
outside of the domain of the contract might affect the state of the
contract. One crucial and fundamental piece of external state that is
relevant to most contracts is the current date and time. It is easy to
see that the state of the contract could be affected by the current
date and time, since many deontic commitments, prohibitions, and
permissions are governed by time. More complex external state, such as
market events, could also affect a contract's state.

\paragraph{Operational}
As mentioned, contracts typically define obligations between parties
to perform some \textit{actions}. If we reuse the example from above:
\textit{Party A must deliver the goods by 12:00 on Friday 4 November
  2018 and Party B must pay Party A \pounds 400 within 7 days of
  receiving the goods.} We can clearly identify two actions that these
parties must carry out, the first being the delivery of some goods and
the second being a payment. It is these actions that we would like to
be able to automate. Before that can happen the actions in a contract
must be identified, computer code must be written to automate their
performance, and we must identify the conditions that must be
satisfied in order for those actions to be allowed to take place.

\subsection{Review of Formalisms}
\subsubsection{Event Condition Action}
Goodchild et al.\cite{goodchild2000business} propose a formalism for
contracts in which obligations are specified as ``policies'' through
actions and associated constraints. These constraints can be absolute
temporal constraints, relative temporal constraints, or conditional
constraints. Operationally, these policies would be defined using
XML,

\subsubsection{Normative Statements}
Boulmakoul and Sall\'{e}\cite{boulmakoul2002integrated} propose
normative statements as a formalism for modelling contracts. Each
normative statement defines some obligation between parties. A
normative statement has the following form:

\[l : f \mapsto D_{i_1,i_2}(a < T) \]

Where $l$ is a label, $f$ is some predicate, $D$ is some deontic
relationship (either obligation, permission or prohibition), $i_1$ and
$i_2$ are identities of parties, $a$ is an action, and $T$ is a
temporal constraint. The statement reads as ``when f holds, $i_1$ is
obliged/permitted/prohibited (depending on D) $i_2$ to achieve/perform
$a$ before $T$''

\subsubsection{Functional Programming}
Peyton Jones and Eber\cite{jones2003write} make use of the
compositional and modular nature of functional programming and use it
to model contracts. They design an algebraic data type to represent a
contract which, in its most basic form, is an atomic obligation
between parties. The definition is enhanced through more complex
binary operators and predicates.

\subsubsection{Finite State Machines}
Molina-Jimenez et al.\cite{molina2004run} propose a formalism that is
similar to Lee's\cite{lee1988} trans assertions which will be
explored later. In this formalism, the life cycle of a contract is
defined as a series of states. Progressions between states occur as a
result of actions. Constrained obligations are represented by having
transitions to a \textit{default} state if some action is not
performed within that constraint (i.e. some absolute temporal
constraint). The system is unique in that the global state of a
contract is not tracked by the machine. Instead, each party has a
respective machine that tracks their progression through the contract
from their perspective.

\subsubsection{Business Contract Language}
Milosevic et al.\cite{linington2004unified}\cite{milosevic2004design}
propose the Business Contract Language (BCL). The language consists of
roles and policies. Policies encapsulate deontic semantics, as well as
associated relative temporal and state based ordering, and absolute
temporal constraints.

\subsubsection{Process Algebra}
Andersen et al\ \cite{andersen2006compositional} propose an approach
that is similar to that of Peyton Jones and Eber \cite{jones2003write}
in the sense that contracts are defined recursively. The base, atomic
definition of a contract is either \textit{success}, signifying that
all obligations have been met, or \textit{failure}, signifying that a
party has defaulted.

\subsubsection{Dynamic Logic}
Prisacariu and Schneider \cite{prisacariu2007formal} propose a
contract language $\mathcal{CL}$. $\mathcal{CL}$ is made up of
elements of deontic, dynamic and temporal logics. An extension to
$\mathcal{CL}$ was proposed by Fenech et al.\cite{fenech2009automatic}
Wtih this extension, they are able to specify obligations, permissions
and prohibitions with state based constraints, and relative temporal
constraints. The system also has support for reparations.

\subsubsection{Defeasible Logic}
Defeasible deontic logic of violation as proposed by Governatori
\cite{governatori2005representing} and later refined by Governatori
and Pham \cite{governatori2009dr} extends defeasible logic
\cite{nute1994defeasible} and adds deontic modalities. In its basic
form, defeasible logic provides a method of reasoning about rules and
precedence among those rules. Rules are similar to implications in
that if we have
\[ r_1 : \alpha \mapsto \beta \]
Then if $\alpha$ is known to be true, $\beta$ is also known to be
true. However, contradictory facts can defeat previously known
facts. For example if we have
\[r_2 : \alpha \mapsto \neg \beta \]
And $r_2 > r_1$, then $r_2$ has precedence over $r_1$ and therefore we
know that $\neg \beta$ must be true. The extension proposed by
Governatori and Pham takes defeasible logic and introduces deontic
predicates.

\subsubsection{Lee's Formalism and Recent Extensions}
Lee \cite{lee1988} proposes a formalism for capturing contract
semantics that brings together many different tools and frameworks
into a single cohesive framework. Pithadia \cite{hirsch} identifies
weaknesses in Lee's formalism and proposes extensions to rectify these
problems. Vanca \cite{vanca2018semantics} does the same, and makes
meaningful contributions that build on Lee's model and Pithadia's
extensions.

\subsection{Evaluation of Reviewed Formalisms} \label{eval}

Hvitved\cite{hvitved2011contract} identifies 16 requirements that a
formalism must support or fulfil to be deemed adequate for modelling
contracts. The requirements are listed below:

\begin{enumerate}[itemsep=0mm]
\item Contract model, contract language, and a formal semantics.
\item Contract participants.
\item (Conditional) commitments.
\item Absolute temporal constraints.
\item Relative temporal constraints.
\item Reparation clauses.
\item Instantaneous and continuous actions
\item Potentially infinite and repetitive
\item Time varying, external dependencies
\item History sensitive commitments
\item In-place expressions
\item Parametrised contracts
\item Isomorphic encoding
\item Run time monitoring
\item Blame assignment
\item Amenability to (compositional) analysis
\end{enumerate}

Hvitved\cite{hvitved2011contract} reviews the formalisms described above and evaluates them
according to his criteria. The results of this analysis are presented
in Table 1.

\begin{table}[H]
\begin{tabular}{|l|l|l|l|l|l|l|l|l|}
\hline
    & Lee & Goo & Bou & Pey & Mol & Mil & And & Pri \\ \hline
R1  &     &     &     &     &     &     & X   & X   \\ \hline
R2  & X   &     & X   &     &     & X   & X   &     \\ \hline
R3  & X   & X   & X   & X   & X   & X   & X   & X   \\ \hline
R4  & X   & X   & X   & X   & (X) & X   & X   &     \\ \hline
R5  & X   &     &     & X   & X   & X   & X   & X   \\ \hline
R6  & X   &     & X   &     & X   & X   & X   & X   \\ \hline
R7  & X   &     &     &     &     &     &     &     \\ \hline
R8  & X   &     & X   & X   &     &     & X   & X   \\ \hline
R9  &     &     &     & X   &     &     & X   &     \\ \hline
R10 &     &     &     &     &     &     & X   &     \\ \hline
R11 &     & X   &     & X   &     &     & X   &     \\ \hline
R12 & X   &     &     & X   &     &     & X   &     \\ \hline
R13 & X   &     & X   & (X) &     & X   &     &     \\ \hline
R14 & X   & X   & X   &     & X   & X   & X   & X   \\ \hline
R15 & (X) &     & (X) &     &     &     &     &     \\ \hline
R16 & X   &     &     & X   & X   & X   & X   & X   \\ \hline
\end{tabular}
\caption{Hvitved’s formalisms comparison matrix. The compared
  formalisms are the following: Lee is Lee's formalism\cite{lee1988},
  Goo is Goodchild et al.\cite{goodchild2000business}, Bou is
  Boulmakoul and Salle\cite{boulmakoul2002integrated}, Pey is Peyton
  Jones and Eber\cite{jones2003write}, Mol is Molina-Jimenez et
  al\cite{molina2004run}, Mil is Milosevic et
  al.\cite{linington2004unified}, And is Andersen et
  al\cite{andersen2006compositional}. and Pri is Prisacariu and
  Schneider\cite{prisacariu2007formal}}
\end{table}

As noted by Vanca \cite{vanca2018semantics}, the formalisms proposed by
Andersen et al. \cite{andersen2006compositional}. and Prisacariu et
al. \cite{prisacariu2007formal} are the only ones that consist of
complete formal models, languages, and semantics. However, there is no
proof that these formalisms can correctly replicate the semantics of
contracts. Additionally, these formalisms fail to meet requirement 7:
support for instantaneous and continuous actions. This would be render
them unsuitable for modelling derivatives contracts as these contracts
typically prescribe some long term, continuous obligations. While it
does not satisfy all of the requirements, Lee's
framework \cite{lee1988} is the next best option.

The law firm King \& Wood Mallesons published a report in
collaboration with ISDA which highlights some further requirements
that are specific to the finance
industry\cite{isdakingwoodmall}. Namely, smart contracts would have to
be constructed with strict adherence to regulatory standards with an
emphasis on safety and transparency, while still meeting the
performance expectations of their users. The report also identifies a
benefit of working in the domain of financial derivatives: the ISDA
Common Domain Model (CDM). The CDM is a dictionary of events and
actions that can occur during the lifecycle of a derivatives
contract. The data is accessible in an unambiguous and machine
interpretable way, and can therefore be integrated into a formal model
for derivatives contracts.

In the next section, different aspects of Lee's framework are
presented and assessed. The analysis is done by looking at how the
formalism deals with the deontic, temporal, and operational aspects of
contracts. Additionally, this analysis looks at how the separability
problem makes it difficult to develop formalisms to exclusively model
each individual aspect.

\section{Review of Lee's Framework}
\subsection{Introduction to the Framework}
Lee \cite{lee1988} presents a combined framework that is designed to
model the various features of contracts identified earlier. Clack
\cite{clack2018smart} finds that the temporal aspects of a contract
are very difficult to separate from the deontic aspects. He calls this
the \textit{separability problem}. Temporal expressions are rarely
found in isolation. Discrete time values in contracts are typically
used to express deadlines for obligations. Likewise, continuous time
intervals are generally used to express the duration of a prohibition
or permission.

The structure of Lee's formalism is influenced by this issue. The
components of the formalism simultaneously express both deontic and
temporal semantics. Additionally, the notion of contract state and
state transitions are also encapsulated by the formalism.

\subsubsection{Summary of the Components of the Formalism}

\textbf{Deontic Logic}: Deontic logic is used to model the
obligations, prohibitions, and permissions that are stipulated in the
contract. Lee's formalism makes use of an extension of the deontic
logic model proposed by Von Wright\cite{von1968essay}.

\noindent\textbf{Petri Nets}: Lee uses Petri Nets to visualise
conditional state changes in the lifecycle of a contract. Changes in
state are triggered by performance or non-performance of
actions. Consequently, the Petri Net can be thought of as a way of
modelling both deontic and relative temporal features of the contract.

\noindent\textbf{Trans Predicate} The trans predicate is a syntactical
representation of the states and transitions of a Petri Net. Lee uses
this syntax in his ``logic programming formulation'' which is an
attempt to implement a working prototype of his formalism.

\noindent\textbf{Rescher Urquhart Calculus}: Lee employs and extends
the Rescher Urquhart calculus to model absolute temporal
constraints. This tool is also somewhat inseparable from deontic
features of the contract, since instances of single discrete time
values or time spans are typically linked to some deontic relationship.

\subsection{Analysis of the Components of Lee's Framework}

\subsubsection{Deontic Logic}
Contracts typically define obligations between two or more transacting
parties. The ISDA Master Agreement defines many such relationships,
where parties agree to fulfil obligations to each other such as
payments or other actions. We require a formal model to be able to
create representations of such relationships.

Deontic logic is a formal model for representing rights and
obligations proposed by Von Wright \cite{von1968essay}. This model
contains an operator to denote obligation:

\[O\phi\]

Meaning that $\phi$ is obligatory. From this further operators are
developed for the notion of permission:

\[P \phi \Leftrightarrow \neg O \neg \phi\]

Meaning that something is permitted if there is no obligation
\textit{not to do it}. And finally for prohibition:

\[F \phi \Leftrightarrow O \neg \phi \]

Meaning that something is prohibited if it is obligatory \textit{not
  to do it}

Lee \cite{lee1988} adopts deontic logic in developing his formal model for
contracts. Lee also adopts a set of axioms to complete the model.

He also introduces an adaptation of the \textit{sanctions} operator:
$S \phi'$ which was first proposed by
Anderson \cite{anderson1956formal}. Lee adopts a relaxed version of
Anderson's model:

\[O \phi \Leftrightarrow ( \neg \phi \Rightarrow S ) \]

This means that for some action to be obligatory, there must exist
some sanctions for non performance of that action. In the real world
the nature of these sanctions may not be defined explicitly in the
contract, and may instead be decided by a court or some form of
arbitration.

This system satisfies requirement 3 as specified by Hvitved. Namely,
it allows us to model the commitments a party has. However, in reality
the deontic aspects in a contract are most often accompanied by some
temporal or conditional constraint. For example, obligations may have
deadlines, prohibitions may apply only for some specified time period,
or some permission may be granted only after some obligation is
fulfilled. Clack and Vanca \cite{clack2018temporal} provide a thorough
report on this subject, identifying many such examples within the ISDA
Master Agreement. The model described above cannot express these links
between deontic elements of the contract and associated discrete
temporal instances, continuous temporal time spans, or conditions.

\subsubsection{Petri Nets, T-Calculus, and the Trans Predicate}
To address the issue describe above, Lee \cite{lee1988} presents a
system for reasoning about sequential orderings of events that may
occur over the life of a legal agreement. Lee uses Petri nets to
visualise the state changes that may occur over the course of a
contract. Through the concept of contracts ``moving forward'' through
states, Petri nets can help to model the \textit{relative} temporal
order of events that may occur. Each state models a set of rights and
obligations, and at each state, there may be a set of states that the
system can progress to. Typically the choice of the next state is
dependent on the performance of some action. In this sense, Petri nets
simultaneously model temporal, operational, and deontic aspects of legal
agreements.

Lee \cite{lee1988} extends Von Wright's \cite{von1968essay} T-Calculus
and employs it as a syntactic representation of Petri Nets. This model
allows us to specify the relative order of contract ``states''. The
key operator is $T$.
\[\phi T \psi \]
means that $\psi$ follows $\phi$. We can also represent choices:
\[\phi T (\psi \vee \theta)\]

Lee \cite{lee1988} extends this concept in his definition of the
$trans$ predicate.  $trans(s1,s2,e)$ has the meaning: ``we transition
from state $s1$ to state $s2$ if event $e$ occurs''. This construct is
equivalent to conditional state transitions in Petri nets.

Vanca \cite{vanca2018semantics} provides an example of the trans predicate being used to model the
following pair of conditional obligations:

\textit{Party A will pay party B a sum of £10,000,000 by 30/08/2018. In
  exchange, B agrees to transfer Building BC to A on 10/09/2018.}

\[
trans([s(0,_0)],[s(1,_1)],A : rb(30−Aug−18) : Pay(B,10 000 000)) \] \[
trans([s(1,_1)],[s(2,_2)],B : rb(10−Sep−18) : Transfer(Building BC))
\]

Notice that this example tackles an issue that has not yet been
discussed: absolute time constraints. This is one of the limitations
of the model as it stands: it can express relative temporal and
conditional relationships between states, but not absolute temporal
constraints. In the next section we will see how Lee models absolute
temporal constraints using the RU Calculus.

Pithadia \cite{hirsch} extends the trans predicate further to express
prohibitions and permissions. Vanca \cite{vanca2018semantics} finds
basic issues with the way these constructs are presented and with the
semantics for the representation of permissions.

\subsubsection{Lee's Adaptation of the RU Calculus}
Lee's \cite{lee1988} framework for modelling absolute temporal
expressions is based on the Rescher and Urquhart temporal logic
system. In this system we have the $R$ operator, with $R_t \phi$
meaning that $\phi$ is realised at time $t$. In addition, a total
ordering of times is introduced, allowing for absolute time
references. Lee expands on this system by introducing notation to
allow for the representation of time spans, and to denote when an some
event is realised during a time span. This is crucial for use in smart
contracts. In Lee's model, this temporal framework is integrated with
deontic logic to represent deadlines (i.e.\ obligations with some
fixed due date) and ongoing prohibitions.

Lee \cite{lee1988} acknowledges some shortcomings of this model. First,
he notes that the model assumes that the base time units used are
equal in length, which is problematic if months or years are desired
as a basis. Second, the system cannot handle continuous time.

Clack and Vanca \cite{clack2018temporal} identify some issues with
Lee's framework. One criticism is that the system cannot handle all of
the complex or nuanced temporal expressions that sometimes arise in
the natural language of legal text. For example: in Lee's framework,
there is no way of expressing ``business days''. Another example: it
is difficult to express the set of dates corresponding to ``the first
Friday of every month'' without directly specifying each day itself.

Vanca \cite{vanca2018semantics} identifies one further issue with the
model: the user is forced to choose some base unit of time for the
system. This could prove problematic in cases where different
contracts that use the same system are dealing with vastly different
timescales. Vanca contrasts the example of a web hosting company that
promises service uptimes with millisecond precision to derivatives
contracts that deal with infrequent payments over many years.

\subsection{Implementation of the Combined Formalism}
Lee \cite{lee1988} describes how his formalism would be implemented in
practice. The key component of this implementation would be the trans
predicate, as well as the syntax that Lee developed to describe
relative and absolute temporal constraints. Obligations would be
modelled as demonstrated in earlier examples. A key feature of Lee's
implementation is modularity and parameterisation. This allows the
smart contract engineer to define reusable components that could be
used in many different smart contract instances.

Earlier in this paper, the operational aspects of contracts were
identified along with the deontic and temporal aspects as being
something that should be modelled by a formalism. Furthermore, the
automation of these operational aspects was identified as a defining
feature of smart contracts. Lee \cite{lee1988} does not go as far as
describing how these operational aspects would be automated in this
formalism. He instead describes a system by which performance of a
contract could be monitored through a kind of simulation environment.

\section{Summary and Future Work}
\subsection{Summary}
This paper initially provided a brief introduction to the problem of
contract formalisation. In doing so, the necessary components of smart
contract formalisms were explained. Then, various different formalisms
were briefly presented, along with an explanation of the criteria that
would be used to evaluate them. Lee's \cite{lee1988} framework was
selected as the best of those presented. This formalism was then
described in depth and analysed with respect to the criteria
identified earlier. Some issues with Lee's framework were identified,
however a formal gap analysis was not conducted.

\subsection{Future Work}
The requirements identified by this paper are informal and do not
necessarily provide full coverage of the ISDA Master Agreement. What
is needed is a formal specification of requirements for
formalims. Once this has been produced, a gap analysis must be
conducted to test whether prospective formalisms meet the
requirements. Once this is done, the gaps that have been identified
must be resolved.

\bibliography{review}

\begin{thebibliography}{10}

\bibitem{lee1988}
R.~M. Lee, ``A logic model for electronic contracting,'' {\em Decision support
  systems}, vol.~4, pp.~27--44, Mar. 1988.

\bibitem{aberdeen2006}
V.~Patel, ``Contract lifecycle management and the cfo: Optimizing revenues and
  capturing savings,'' {\em Aberdeen-Group, Boston, NY}, Apr. 2007.
\newblock
  https://images.treasuryandrisk.com/treasuryandrisk/historical/SiteCollectionDocuments/aberdeen-contract-lifecycle.pdf/.

\bibitem{isdainfra2016}
ISDA, ``The future of derivatives processing and market infrastructure,'' {\em
  ISDA Whitepaper}, Sept. 2016.
\newblock https://www.isda.org/a/UEKDE/infrastructure-white-paper.pdf.

\bibitem{clack2016smart-foundations}
C.~D. Clack, V.~A. Bakshi, and L.~Braine, ``Smart contract templates:
  foundations, design landscape and research directions,'' {\em CoRR},
  vol.~abs/1608.00771, 2016.

\bibitem{DBLP:journals/corr/ClackBB16a}
C.~D. Clack, V.~A. Bakshi, and L.~Braine, ``Smart contract templates: essential
  requirements and design options,'' {\em CoRR}, vol.~abs/1612.04496, 2016.

\bibitem{hvitved2011contract}
T.~Hvitved, {\em Contract Formalisation and Modular Implementation of
  Domain-Specific Languages}.
\newblock PhD thesis, The Faculty of Science, University of Copenhagen, 2011.

\bibitem{goodchild2000business}
A.~Goodchild, C.~Herring, and Z.~Milosevic, ``Business contracts for b2b.,''
  {\em ISDO}, vol.~30, 2000.

\bibitem{boulmakoul2002integrated}
A.~Boulmakoul and M.~Sall{\'e}, ``Integrated contract management,'' in {\em
  Proceedings of the 9th Workshop of the HP OpenView University Association},
  2002.

\bibitem{jones2003write}
S.~P. Jones and J.-M. Eber, ``How to write a financial contract,'' {\em
  Citeseer}, 2003.

\bibitem{molina2004run}
C.~Molina-Jimenez, S.~Shrivastava, E.~Solaiman, and J.~Warne, ``Run-time
  monitoring and enforcement of electronic contracts,'' {\em Electronic
  Commerce Research and Applications}, vol.~3, no.~2, pp.~108--125, 2004.

\bibitem{linington2004unified}
P.~F. Linington, Z.~Milosevic, J.~Cole, S.~Gibson, S.~Kulkarni, and S.~Neal,
  ``A unified behavioural model and a contract language for extended
  enterprise,'' {\em Data \& Knowledge Engineering}, vol.~51, no.~1, pp.~5--29,
  2004.

\bibitem{milosevic2004design}
Z.~Milosevic, S.~Gibson, P.~F. Linington, J.~Cole, and S.~Kulkarni, ``On design
  and implementation of a contract monitoring facility,'' in {\em Electronic
  Contracting, 2004. Proceedings. First IEEE International Workshop on},
  pp.~62--70, IEEE, 2004.

\bibitem{andersen2006compositional}
J.~Andersen, E.~Elsborg, F.~Henglein, J.~G. Simonsen, and C.~Stefansen,
  ``Compositional specification of commercial contracts,'' {\em International
  Journal on Software Tools for Technology Transfer}, vol.~8, no.~6,
  pp.~485--516, 2006.

\bibitem{prisacariu2007formal}
C.~Prisacariu and G.~Schneider, ``A formal language for electronic contracts,''
  in {\em FMOODS}, vol.~7, pp.~174--189, Springer, 2007.

\bibitem{fenech2009automatic}
S.~Fenech, G.~J. Pace, and G.~Schneider, ``Automatic conflict detection on
  contracts,'' in {\em International Colloquium on Theoretical Aspects of
  Computing}, pp.~200--214, Springer, 2009.

\bibitem{governatori2005representing}
G.~Governatori, ``Representing business contracts in ruleml,'' {\em
  International Journal of Cooperative Information Systems}, vol.~14,
  no.~02n03, pp.~181--216.

\bibitem{governatori2009dr}
G.~Governatori and D.~H. Pham, ``Dr-contract: An architecture for e-contracts
  in defeasible logic,'' {\em International Journal of Business Process
  Integration and Management}, vol.~4, no.~3, pp.~187--199, 2009.

\bibitem{nute1994defeasible}
D.~Nute, ``Defeasible logic,'' in {\em Handbook of logic in artificial
  intelligence and logic programming (vol. 3)}, pp.~353--395, Oxford University
  Press, Inc., 1994.

\bibitem{hirsch}
H.~J. Pithadia, ``{Capturing Language Semantics of Smart Contracts},'' Master's
  thesis, Departament of Computer Science, UCL, Sep 2016.

\bibitem{vanca2018semantics}
G.~Vanca, ``The semantics of smart contracts used in banking and financial
  services,'' Master's thesis, Departament of Computer Science, UCL, Apr. 2018.

\bibitem{isdakingwoodmall}
{ISDA and King \& Wood Mallesons}, ``Smart derivatives contracts: From concept
  to construction,'' {\em ISDA Whitepaper}, Oct. 2018.
\newblock
  https://www.isda.org/a/cHvEE/Smart-Derivatives-Contracts-From-Concept-to-Construction-Oct-2018.pdf.

\bibitem{clack2018smart}
C.~D. Clack, ``Smart contract templates: Legal semantics and code validation,''
  {\em Journal of Digital Banking}, vol.~2, no.~4, pp.~338--352, 2018.

\bibitem{von1968essay}
G.~H. Von~Wright, ``An essay in deontic logic and the general theory of
  action,'' {\em Acta philosophica Fennica}, vol.~21, 1968.

\bibitem{anderson1956formal}
A.~R. Anderson, ``The formal analysis of normative systems,'' tech. rep., Yale
  University Interaction Lab, New Haven CT, 1956.

\bibitem{clack2018temporal}
C.~D. Clack and G.~Vanca, ``Temporal aspects of smart contracts for financial
  derivatives,'' {\em CoRR}, vol.~abs/1805.11677, 2018.

\end{thebibliography}
\bibliographystyle{ieeetr}
\end{document}